\documentclass[preprint,aps,floats,subeqn,showpacs]{revtex4}
\usepackage{epsfig}

\def\real{I\negthinspace R}

\newcommand{\be}{\begin{equation}}
\newcommand{\ee}{\end{equation}}
\newcommand{\bea}{\begin{eqnarray}}
\newcommand{\eea}{\end{eqnarray}}
\newcommand{\bml}{\begin{mathletters}}
\newcommand{\eml}{\end{mathletters}}

\begin{document}

\tighten

\preprint{IUB-TH-041}
\draft




\title{ Spherically symmetric Yang-Mills solutions  in
a $(4+n)$- dimensional space-time}
\renewcommand{\thefootnote}{\fnsymbol{footnote}}
\author{ Yves Brihaye\footnote{Yves.Brihaye@umh.ac.be}}
\affiliation{Facult\'e des Sciences, Universit\'e de Mons-Hainaut,
7000 Mons, Belgium}
\author{Fabien Clement}
\affiliation{Facult\'e des Sciences, Universit\'e de Mons-Hainaut,
7000 Mons, Belgium}
\author{Betti Hartmann\footnote{b.hartmann@iu-bremen.de}}
\affiliation{School of Engineering and Sciences, International University Bremen (IUB),
28725 Bremen, Germany}

\date{\today}
\setlength{\footnotesep}{0.5\footnotesep}

\begin{abstract}
We consider the Einstein-Yang-Mills Lagrangian
in a $(4+n)$-dimensional space-time.
Assuming the matter and metric fields to be 
independent of the $n$
extra coordinates, a spherical symmetric Ansatz for
the fields leads to a set of
coupled ordinary differential equations. We find that
for $n > 1$ only solutions with either  one non-zero Higgs field or
with all Higgs fields constant exist.
We construct the
analytic solutions which fulfill this conditions
for arbitrary $n$, namely the Einstein-Maxwell-dilaton solutions.

We also present generic solutions of the
effective
4-dimensional Einstein-Yang-Mills-Higgs-dilaton model, which
possesses $n$ Higgs triplets coupled in a specific way to
$n$ independent dilaton fields. These solutions are the abelian Einstein-Maxwell-
dilaton solutions and analytic non-abelian solutions, which have
diverging Higgs fields. In addition, we construct numerically 
asymptotically flat and finite energy solutions for $n=2$. 
\end{abstract}

\maketitle

\section{Introduction}
In an attempt to unify electrodynamics and general relativity,
Kaluza introduced an extra - a fifth dimension \cite{kaluza} and assumed all
fields to be independent of the extra dimension.
Klein \cite{klein} followed this idea, however, he assumed the fifth dimension
to be compactified on a circle of Planck length. The resulting theory
describes $4$-dimensional Einstein gravity plus Maxwell's equations. 
One of the new fields appearing in this model
is the dilaton, a scalar companion of the metric tensor.
In an analogue way, this field arises in the
low energy effective action of superstring theories and is
associated with the classical scale invariance of these models \cite{maeda}.

When studying spherically symmetric solutions in higher dimensional systems,
two possible approaches seem possible: a) to assume the solutions
to be spherically symmetric in the full $d$ dimensions or b) to assume
the solutions to be spherically symmetric only in 4 dimensions. The solutions
obtained in the first approach are surely important at very high energies, i.e.
very early stages of the universe. The second approach, which assumes the extra dimensions
to be ``spectator'' is of importance for physics in the universe today.

Volkov recently followed the idea of extra dimensions and
constructed a $(4+1)$ dimensional Einstein-Yang-Mills
(EYM) system \cite{volkov}. It turned out that the EYM particles
are completely destroyed by gravity, but that, however, particle-like solutions
(so-called ``EYM vortices'') exist if one assumes the fields to be independent of the extra coordinate.
The system then reduces to an effective $4$-dimensional
Einstein-Yang-Mills-Higgs-dilaton (EYMHD) model, which was studied in
detail in \cite{bh1}. Through the dimensional reduction, one Higgs triplet
and a dilaton field appears in the model.

In this paper, we study the system of \cite{volkov} in $(4+n)$ dimensions, assuming all
fields to be independent of the $n$ extra coordinates.
We give the model and equations of motion in Section II. The solutions are
spherically symmetric in 4 dimensions, while the extra dimensions
(which are ``spectator'') are associated with a Ricci flat manifold.
It turns out that in $n > 1$ dimensions only solutions
with one non-zero Higgs field or with all Higgs fields constant exist. 
We give the analytic solutions which fulfill this condition 
for generic $n$, namely the abelian Einstein-Maxwell-dilaton (EMD).
In Section III, we present the 4-dimensional
effective Einstein-Yang-Mills-Higgs-dilaton (EYMHD) model, which has
$n$ independent Higgs and dilaton fields. Since in this model, no constraint
on the Higgs fields arises, we construct generic solutions.
We give the analytic solutions available, namely the abelian 
Einstein-Maxwell-dilaton (EMD) solutions and ``non-abelian'' solutions with
diverging Higgs fields.
Finally, we present our numerical solutions of the effective
model for $n=2$.
We give our conclusions in Section IV.

\section{The $(4+n)$-dimensional Einstein-Yang-Mills model}

The Einstein-Yang-Mills Lagrangian
in $d=(4+n)$ dimensions is
given by:

\begin{equation}
\label{action}
  S = \int \Biggl(
    \frac{1}{16 \pi G_{(4+n)}} R   - \frac{1}{4 e^2}F^a_{M N}F^{a M N}
  \Biggr) \sqrt{g^{(4+n)}} d^{(4+n)} x
\end{equation}
with the SU(2) Yang-Mills field strengths
$F^a_{M N} = \partial_M A^a_N -
 \partial_N A^a_M + \epsilon_{a b c}  A^b_M A^c_N$
, the gauge index
 $a=1,2,3$  and the space-time index
 $M=0,...,(4+n)-1$. $G_{(4+n)}$ and $e$ denote
respectively the $(4+n)$-dimensional Newton's constant and the coupling
constant of the gauge field theory. $G_{(4+n)}$ is related to the Planck mass
$M_{pl}$ by $G_{(4+n)}=M_{pl}^{-(2+n)}$ and $e^2$ has the dimension of $[{\rm length}]^n$.
In the following, we denote the coordinates $x_{(3+k)}$ by $y_k$ with 
$k=1,...,n$.

If both the matter functions and the metric functions
are independent on $y_k$, the fields can be
parametrized as follows:
\begin{equation}
g^{(4+n)}_{MN}dx^M dx^N = 
e^{-\Xi}g^{(4)}_{\mu\nu}dx^{\mu}dx^{\nu}
+\sum_{k=1}^n e^{2\zeta_k} (dy^k)^2 
\ , \ \mu , \nu=0, 1, 2, 3 
\end{equation}
with
\begin{equation}
\Xi=\sum_{k=1}^n \zeta_k 
\end{equation}
and
\begin{equation}
A_M^{a}dx^M=A_{\mu}^a dx^{\mu}+ \sum_{k=1}^n \Phi_k^a dy^k   \  . \
\end{equation}
$g^{(4)}$ is the $4$-dimensional metric tensor 
and the $\zeta_{j}$ and  $\Phi_j^a$, $j=1,...,n$,  
play the
role of  dilatons and Higgs fields, respectively. 
 
Note that in the case $n=1$, the above parametrization coincides
with the one in \cite{volkov}.

\subsection{Spherically symmetric Ansatz}

For the metric the spherically symmetric Ansatz
in Schwarzschild-like coordinates reads \cite{weinberg}:
\begin{equation}
ds^{2}=g^{(4)}_{\mu\nu}dx^{\mu}dx^{\nu}=
-A^{2}(r)N(r)dt^2+N^{-1}(r)dr^2+r^2 d\theta^2+r^2\sin^2\theta
d^2\varphi
\label{metric}
\ , \end{equation}
with
\begin{equation}
N(r)=1-\frac{2m(r)}{r}
\ . \end{equation}
In these coordinates, $m(\infty)$ denotes the (dimensionful) mass of
the field configuration.\

For the gauge and Higgs fields, we use the purely magnetic hedgehog ansatz
\cite{thooft} :
\begin{equation}
{A_r}^a={A_t}^a=0
\ , \end{equation}
\begin{equation}
{A_{\theta}}^a= (1-K(r)) {e_{\varphi}}^a
\ , \ \ \ \
{A_{\varphi}}^a=- (1-K(r))\sin\theta {e_{\theta}}^a
\ , \end{equation}
\begin{equation}
\label{higgsansatz}
{\Phi}^a_{j}=c_j v H_j(r) {e_r}^a \ \ , \ \ j=1,...,n \ ,
\end{equation}
where $v$ is a mass scale, while $c_j$ are dimensionless constants
determining the vacuum expectation values of the Higgs fields 
$\langle \Phi_j \rangle = c_j v$.
In absence of a Higgs potential these have to be set by hand.
Finally, the dilatons are scalar fields depending only on $r$~:
\begin{equation}
\zeta_j =\zeta_j(r) \ \ , \ \ k = 1,...,n 
\ . \end{equation}

With these conventions, the non-vanishing components of the 
$(4+n)$-dimensional
energy momentum tensor read:
\begin{eqnarray}
& T_0^0 &= - e^{\Xi}
\left(\sum_{k=1}^n (A_k  + C_k ) + B + D\right) \ , \nonumber \\
& T_1^1 &= - e^{\Xi}
\left(\sum_{k=1}^n (-A_k  + C_k ) - B + D\right) \ , \nonumber \\
& T_2^2 &= T_3^3 = - e^{\Xi}
\left(\sum_{k=1}^n A_k  - D\right) \ ,  \nonumber \\
& T_j^j &= - e^{\Xi}
\left(-2 (A_j+C_j) 
+ \sum_{k=1}^n (A_k + C_k) + B + D\right) \ , \ j=1,...,n \ ,
\nonumber \\
& T_j^k &=  e^{\Xi} e^{-(\zeta_j + \zeta_k)} \left(\frac{N}{2} H'_j H'_k
                                       + \frac{K^2}{x^2} H_j H_k\right)
\ \ , \ \ j \neq k  \ \ , \
\end{eqnarray}
 where we use the abbreviations
 \be
 A_j = \frac{1}{2} e^{-2 \zeta_j} N (H_j')^2 \ \ , \ \
 C_j = e^{-2 \zeta_j} K^2 H_j^2 \frac{1}{x^2} \ \  \ \
 \ee
 and
 \be
  B =  e^{\Xi} \frac{1}{x^2} N (K')^2 \ \ , \ \
  D =  e^{\Xi} \frac{1}{2 x^4}  (K^2 - 1)^2
 \ee
 The radial variable  $x=e v r$ was introduced and
 the prime denotes the  derivative with respect to $x$. 

\subsection{Equations of motion}

Defining the coupling constant $\alpha = v \sqrt{G_{(4+n)}}$ and the 
mass function $\mu=evm$ 
we obtain two first order equations for the two metric functions
\be
\label{mu}
\mu' = \alpha^2 x^2\left(\sum_{k=1}^n (A_k+C_k) 
+ B + D\right) + \frac{1}{2}N x^2 \Theta  \ ,
\ee
\be
\label{A}
   A' = \alpha^2 A x \left(\sum_{k=1}^n e^{-2\zeta_k} (H_k')^2 
   + 2 e^{\Xi} \frac{K'^2}{x^2}\right) + x A \Theta
\ee
with 
\be
  \Theta \equiv \frac{3}{4}\left(\sum_{k=1}^n (\zeta_k')^2 
+ \frac{2}{3} \sum_{k > k'} \zeta_{k}' \zeta_{k'}' \right)  \ .
\ee
The components of the Einstein equations related to the extra dimensions lead
(after suitable linear combinations) to equations for the
$n$ dilaton fields~:
\begin{equation}
\label{zeta}
(x^2 A N \zeta_j ')'
= \alpha^2 A x^2 e^{-\Xi}
\left[ \frac{2}{n+2}\sum_{M=0}^{d-1} T_M^M -2 T_j^j\right]  \  \ , \ \ j=1,...,n \ .
\end{equation}
The field equations for the gauge field and the
$n$ Higgs fields read respectively
\be
\label{k}
   (e^{\Xi}ANK')'
   = A\left( \frac{1}{x^2}e^{\Xi}K(K^2-1)
       + \sum_{k=1}^n e^{-2 \zeta_k} K H_k^2\right)  \ ,
\ee
\be
\label{h}
   (e^{-2\zeta_j}x^2 A N H_j')' = 2 A e^{-2 \zeta_j} K^2 H_j \ \ , \ \
   j=1,...,n   \ .
\ee
Finally, since the off-diagonal components of the
Einstein tenor vanish, we obtain an extra constraint on the 
fields from the $jk$-components of the energy-momentum tensor:
\begin{equation}
\label{constraint}
e^{\Xi} e^{-(\zeta_j + \zeta_k)} \left(\frac{N}{2} H'_j H'_k
                                       + \frac{K^2}{x^2} H_j H_k\right)=0
\end{equation}
Clearly, these equations are symmetric under the simultaneous
exchange $\zeta_{j'},  H_{j'} \leftrightarrow \zeta_j, H_j$. Further, they
are invariant under the rescaling 
\be
   H_{j} \rightarrow {\lambda}^{-1} H_j \ \ , \ \
   x \rightarrow \lambda x \ \ , \ \
   \alpha \rightarrow \lambda^2 \alpha  \ , \ \ \ \lambda \in \real  \ .
\ee

The constraint (\ref{constraint}) will only be fulfilled
for specific cases. 
With our parametrisation, there are two possibilities: 
(i) only {\it one} non-zero Higgs field and arbitrary gauge field $K(x)$
or (ii) constant Higgs fields and $K(x)=0$.

Since we are interested in generic solutions,
we go back to the 4-dimensional effective action motivated by this model
in which, of course, off-diagonal terms of the energy-momentum tensor
don't appear. Before we discuss this effective model, we give the
solutions with constant Higgs fields available in the ``full'' model, namely
the Einstein-Maxwell-dilaton (EMD) solutions.

Let us remark that there are several possibilities of modification
of the ``full'' model considered here, 
which could eventually lead to solutions with generic Higgs and gauge fields, 
namely we could

 (i) add a non-diagonal term of the form $f_{56}dx_5 dx_6$
in the metric, this leads to one extra equation for the $f_{56}$
function, but it has to be checked that this function stays regular,

(ii) start with an SU(N) gauge group and choose the different Higgs
fields in orthogonal SU(2) subalgebras of the Lie algebra of SU(N).

\subsection{Einstein-Maxwell-Dilaton (EMD) solutions}
Assuming the matter fields to be constant:
\begin{equation}
K(x)=0 \ , \ H_j(x)=c_j \ , \ j=1,...,n
\end{equation}
and 
\begin{equation}
\zeta_j(x)=\zeta(x) \ \ \ \forall \  j  \ ,
\end{equation}
we find that the above equations (\ref{mu}), (\ref{A}), (\ref{zeta})
admit exact solutions which are related to the Einstein-Maxwell-Dilaton (EMD) solutions 
\cite{maeda}. Here, we will only discuss the extremal case, which corresponds to
a solution with horizon at the origin. 

The extremal EMD solutions have unit magnetic charge and mass
\begin{equation}
\label{massEMD}
\frac{\mu_{\infty}}{\alpha^2}=\frac{1}{\alpha} \sqrt{\frac{n+2}{2(n+1)}} \ .
\end{equation}
The value of the metric component $N(x)$ at the origin $x=0$ reads:
\begin{equation}
\label{NEMD}
N_{\small EMD}(0)=\left(\frac{n}{2(n+1)}\right)^2
\end{equation}
and the dilaton field is:
\begin{equation}
\zeta_{\small EMD}(x)=\frac{1}{n+1} \ln \left(1-\frac{X_-}{X}\right)
\end{equation}
with
\begin{equation}
X_-=\left(\frac{2(n+1)}{n+2}\right)^{1/4} \ \ , \ \ \frac{x}{\alpha}=X
\left(1-\frac{X_-}{X}\right)^{n/(2n+2)}  \ .
\end{equation}

\section{The effective $4$-dimensional Einstein-Yang-Mills-Higgs-Dilaton (EYMHD) model}
As in the 5-dimensional case \cite{volkov} (in our notation $n=1$) the 
equations given in the previous section (apart from the constraint
(\ref{constraint})) can equally well be derived 
from an effective 4-dimensional
Einstein-Yang-Mills-Higgs-Dilaton (EYMHD) Lagrangian. 
The Lagrangian density for the matter fields then reads:
\begin{eqnarray}
\label{effeL}
   L_M =
&-& \frac{1}{4} e^{2 \kappa \Gamma }F^a_{\mu \nu} F^{a, \mu \nu}
- \sum_{k=1}^n 
\frac{1}{2} e^{-4 \kappa \Psi_k} D_{\mu} \Phi_1^a D^{\mu} \Phi_1^a
\nonumber \\
&-& \frac{1}{2} \left(\sum_{k=1}^n \partial_{\mu} \Psi_k  \partial^{\mu} \Psi_k               
        +\frac{2}{3} \sum_{k > k'} \partial_{\mu} \Psi_k  \partial^{\mu} 
\Psi_{k'}\right)
\end{eqnarray}
with
\begin{equation}
\Gamma=\sum_{k=1}^n \Psi_k  \ .
\end{equation}
The kinetic part in the dilaton fields could be diagonalized, however,
we find it more convenient to leave it in the form above which reveals
the symmetry $\Psi_j, \Phi_j \leftrightarrow \Psi_{j'}, \Phi_{j'}$.

The Lagrangian (\ref{effeL}) is then coupled minimally to Einstein gravity
according to the full action
\begin{eqnarray}
\label{action4}
              S &=& S_G + S_M \nonumber \\
              &=& \int \sqrt{-g^{(4)}} \left(L_G + L_M \right)  d^4 x
\end{eqnarray}
where $L_G = R/(16 \pi G_4)$, $R$ is the Ricci scalar
and $G_4$ is the $4$-dimensional Newton's constant.

Note that the dilaton fields are coupled by an independent coupling
constant $\kappa$ to the gauge and Higgs fields. In this respect, the
dilatons here are treated as independent scalar fields, while in
the action (\ref{action}) they appear as parts of the metric tensor.
 
After the rescaling
\begin{equation}
          \Psi_j  =  v \psi_j \ \ , \ \ \kappa = \frac{\gamma}{v} \ \ , \ \
\alpha=v\sqrt{G_4}
\end{equation}
the resulting set of equations only depends on the coupling constants
$\alpha$ and $\gamma$. We refrain from giving the explicit form of the
equations here, but refer the reader to \cite{bh1} for the case $n=1$.

Note that the equations (\ref{mu}), (\ref{A}), (\ref{zeta}), (\ref{k}), (\ref{h})
become equivalent to the field equations associated to (\ref{action4})
by using the same Ans\"atze for the $4$-dimensional metric, the gauge and
Higgs fields, but by identifying 
\begin{equation}
\zeta_j=2\gamma \psi_j=2\kappa \Psi_j \ \ , \ \ \alpha^2=3\gamma^2  \ . 
\end{equation} 
Remarkably, this identification turns out to be independent on $n$.

This model now has solutions for which 
all $n$ Higgs fields can be non-constant. First, we will present the
4-dimensional effective
counterparts of the abelian Einstein-Maxwell-dilaton solutions. Then, we will
present analytic non-abelian solutions with non-constant but
diverging Higgs fields, which are not available in the full model
since they don't fulfill the constraint (\ref{constraint}). Finally, we will give
the asymptotically flat, finite energy solutions for $n=2$, which we construct
numerically.

\subsection{Einstein-Maxwell-Dilaton (EMD) solutions}

Assuming the matter fields to be constant:
\begin{equation}
K(x)=0 \ , \ H_k(x)=c_k \ , \ k=1,...,n
\end{equation}
and 
\begin{equation}
\psi_k(x)=\psi(x) \ \ \ \forall \  k   \ ,
\end{equation}
we find that the equations associated to the effective action (\ref{action4})
admit exact solutions which are related to the Einstein-Maxwell-Dilaton (EMD) solutions 
\cite{maeda}. As in the previous section, 
we will only discuss the extremal case. Note that now, of course, the solution
depends on both $\alpha$ and $\gamma$. 

The extremal EMD have unit magnetic charge and mass
\begin{equation}
\frac{\mu_{\infty}}{\alpha^2}=
\frac{1}{\sqrt{\alpha^2 + \tilde{\gamma}^2}} \ , 
\end{equation}
where 
\begin{equation}
\tilde{\gamma}^2=\frac{3n}{2+n} \gamma^2  \ .
\end{equation}
The value of the metric component $N(x)$ at the origin $x=0$ reads:
\begin{equation}
N_{\small EMD}(0)=\left(\frac{\tilde{\gamma}^2}{\alpha^2 + \tilde{\gamma}^2}
\right)^2
\end{equation}
and the dilaton field is:
\begin{equation}
\psi_{\small EMD}(x)=\left(\frac{3}{2+n}\right)\frac{1}{\alpha^2 + \tilde{\gamma}^2}  
\ln \left(1-\frac{X_-}{X}\right)
\end{equation}
with
\begin{equation}
X_-=\left(\frac{\alpha^2+\tilde{\gamma}^2}{\alpha^2
}\right)^{1/4} \ \ , \ \ \frac{x}{\alpha}=X  
\left(1-\frac{X_-}{X}\right)^{\tilde{\gamma}^2/(\alpha^2+\tilde{\gamma}^2) }
\end{equation}

\subsection{Analytic, non-abelian solutions}

Very similar to the $n=1$ case \cite{volkov} non-abelian, 
analytic solutions exist in this system. If we assume the limit
$\alpha^2=3\gamma^2$ (which, of course, is the most interesting one),
we find solutions of the form:
\begin{eqnarray}
\label{ads}
& & N(x)=c_1 \ \ , \ \ A = x^{(1+c_2)} \ \ , \ \  \psi_1(x)=\psi_2(x)
=...\equiv\psi(x)=\frac{4\alpha}{\sqrt{3}n} \ln\left(\frac{x}{c_3}\right) \nonumber \\
&  & H_1(x)=H_2(x)=...=H(x)=\sqrt{c_4} \left(\frac{x}{c_3}\right)^{2/n} 
\ \ \ , \ \  K(x)=\sqrt{q}
\end{eqnarray}
where $c_1$, $c_2$, $c_3$ and $c_4$ are constants depending only on $n$, $q$
and $\alpha$
and are given by:
\begin{eqnarray}
& &c_1= \frac{-2q^2n^3(n + 1) + qn^3(2n^2 + 5n + 6) - n^4}{8
qn(n + 1) + 4(2-n)}\ \ , 
\ \ c_2= \frac{2}{n}(1+2\alpha^2 c_4)\ \ , \nonumber \\
& & c_3^2=\frac{ - 4 q^3 \alpha^2 (n + 1) 
+ 2 q^2 \alpha^2 (2 n^2 + 7n + 8) + 4q
\alpha^2( - n^2 - 3n - 3) + 2\alpha^2 n}{2 q^2 n( - n^2 - 3n - 2) +
q (2 n^4 + 5 n^3 + 6 n^2 + 2n - 4) - n^3} \ \ , \nonumber \\
& & c_4  = \frac{1-q}{n c^2_3}  \ , 
\end{eqnarray}
and $q$ satisfies the equation
\begin{equation}
q^3(2n^4 + 4n^3 + 2n^2) + q^2(- 2  n^5 - 3 n^4 
+ n^3 + 10 n^2 + 8  n) + q( 3 n^4 + 2  n^3 +
 3  n^2 + 6 n + 8) - n^3 + n^2 - 2n = 0 \ .
\end{equation}
Solving this condition for $q$ numerically for various
values of $n$, we find one real positive solution for $n=1$, $2$, $3$ and
three positive solutions for $n=4$, $5$, $6$, $7$. However, when
more than one solution for $q$ are available, it turns out that only one of
the solutions is acceptable since for the others some of the remaining
parameters (e.g. $c_4$ or $c_3$) become negative.
In Table 1, we give the acceptable solutions and their numerical values
for the parameters $c_1$, $c_2$, $c_3$, $c_4$:
\begin{center}
{\bf Table 1}: Values of $q$ and constants 
$c_1$, $c_2$, $c_3$, $c_4$ for different $n$
\\
\begin{tabular}{|l|l|l|l|l|l|}
\hline
$n$ & $K(x)=\sqrt{q}$ & $c_1$ & $c_2$  & $c_3^2/\alpha^2$
&$\alpha^2 c_4$   \\
\hline 
$1$    & $0.293$  & $0.016$  & $5.246$ & $1.126$ & $0.811$\\
$2$    & $0.294$  & $0.064$  & $1.702$  & $1.301$ & $0.351$\\
$3$    & $0.280$  & $0.102$  & $0.982$  & $1.297$ & $0.237$\\
$4$    & no real solution & & & &\\
$5$    & $0.249$  & $0.149$  & $0.495$  & $1.581$ & $0.118$\\
$6$    & $0.236$  & $0.151$  & $0.554$  & $0.476$ & $0.331$\\
$7$    & $0.225$  & $0.086$  & $2.402$  & $0.036$ & $3.704$ \\
\hline
\end{tabular}
\end{center}

\subsection{Numerical results for $n=2$}
The $4$-dimensional
effective theory for $n=1$ was studied in \cite{bh1}.

Here, we will put the emphasis on the case $n=2$ and again
discuss the solutions in the limit $\alpha^2=3\gamma^2$.
In this model, two Higgs 
fields naturally occur. Theories involving two Higgs fields are also
interesting from the viewpoint of the supersymmetric extension of
the Standard model.

The case $n=2$ involves two parameters
only, namely $\alpha$ and $\rho\equiv c_2/c_1$.  
As discussed previously in this paper, it is sufficient to consider
$\rho \geq 1$. 

The solutions can then be characterized by their mass~:
\be
   M = \frac{1}{\sqrt{c_1^2 + c_2^2}}\frac{\mu(\infty)}{\alpha^2}  \ ,
\ee
where the first factor is extracted in such a way that the mass of the
solution in the $\alpha=0$ limit is normalized to $1$, corresponding to the BPS 
monopole.

The values $\psi_{1,2}(0)$ will appear to be useful 
to further characterize the solutions. 
The deviation of the solutions from the flat space solutions can be
``measured'' by the values of $A(x)$ at the origin, $A(0)$,
 and by the minimum of $N(x)$, say $N_m$. Since $A(x)$ always increases
monotonically, $A(0)$ also represents the minimum of the function.

In the following, we will discuss our numerical results \cite{foot1} for
two different and fixed values of $\rho$ and varying $\alpha$.

We again discuss the solutions in the limit of $\alpha^2=3\gamma^2$.

\subsubsection{Boundary conditions}

We will study globally regular, asymptotically flat 
solutions of the system above. This implies the following boundary conditions:
\begin{equation}
K(0)=1 \ , \ \ H_j(0)=0 \ , \ \ \partial_{x}\psi_j|_{x=0}=0 \ ,
\ \ \mu(0)=0
\  \label{bc1}
\end{equation}
at the origin and 
 \begin{equation}
K(\infty)=0 \ , \ \ H_j(\infty)=c_j \ ,
\ \ \psi_j(\infty)=0 \ , \ \ A(\infty)=1
\ , \label{bc2} \end{equation}
at infinity. We assume $c_j\neq 0$ in order for the solutions to have
a magnetic charge.
Using the symmetries discussed above, we see that we can
set ${\rm min} \{c_j, j=1,...,n\} = c_1 = 1$ without loosing generality
 and study
the equations with varying $\alpha$ and $c_2, c_3, ..., c_n$ $ > 1$.
Of course, in the case $c_1=c_2=...=c_n = 1$ the equations are symmetric
and we expect $H_1(x) = H_2(x)=...H_n(x)\equiv H(x)$ , 
$\psi_1(x) = \psi_2(x)=...=\psi_n(x)\equiv \psi(x)$.

\subsubsection{$\rho=1$}

In this case, the two Higgs functions $H_1(x)$ and $H_2(x)$
are equal. This, of course, implies immediately that also 
$\psi_1(x)=\psi_2(x)$.

In the flat limit ($\alpha =0$) the solution is the  BPS monopole \cite{ps,bogo}
with $A(x)=N(x)=1$, $\psi_1(x) = \psi_2(x) = 0$ and the 
matter functions $K(x)$, $H_1(x)= H_2(x)$ have the well 
known BPS profiles.

Increasing $\alpha$, our numerical results reveal
that the  solution gets progressively deformed by gravity.
The solutions form a branch on which the
mass $M$ dimnishes with increasing $\alpha$.
The same holds true for the values $A(0)$, $N_m$.

The function $N(x)$ indeed develops a minimum which becomes deeper
while gravity increases. At the same time the dilaton functions
$\psi_{1,2}(x)$ are non trivial for $\alpha\neq 0$. The value of $\psi_{1,2}(x=0)$
is negative and decreases with increasing $\alpha$.

\begin{figure}
\centering
\epsfysize=20cm
\mbox{\epsffile{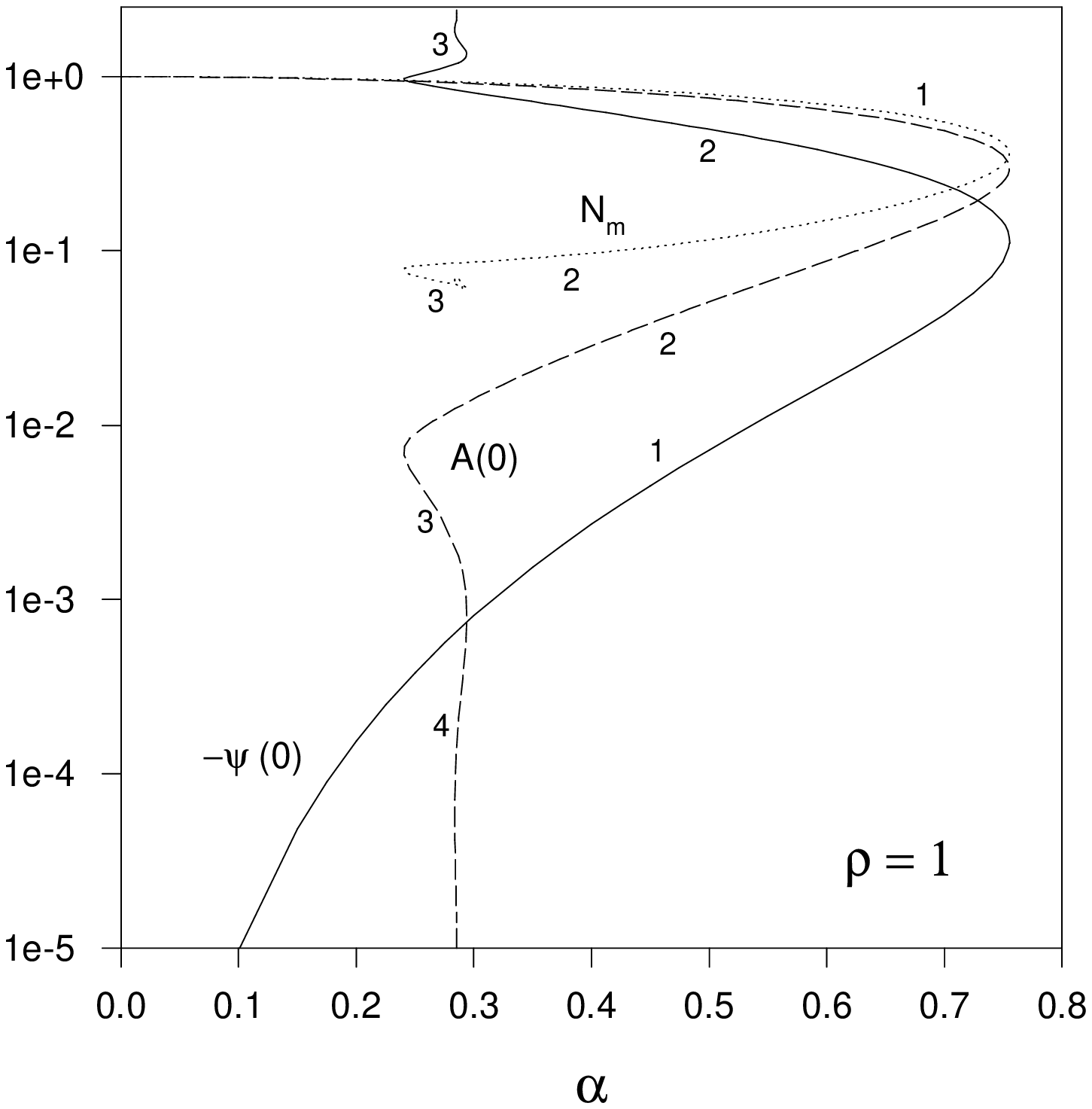}}
\caption{\label{fig1} The values $A(0)$, $N_m$, $-\psi(0)$ are shown
as functions of $\alpha$ for $\rho=1$. The indices ``1'',
``2'', ``3'', ``4'', respectively correspond to the 1., 2., 3. and 4. 
branch of solutions (see also Fig.~\ref{fig2}). }
\end{figure}

Our results are illustrated in Fig.~\ref{fig1}.
The figure also demonstrates that the branch of gravitating solutions
does not exist for arbitrarily large values of the parameter $\alpha$.
Indeed, we find that solutions cease to exist for $\alpha > \alpha^{(1)}_{max}$
and $\alpha^{(1)}_{max}$ thus constitutes a maximal value of $\alpha$. We find that
this critical value depends on the parameter $\rho$ and is given here by:
 \be
       \alpha^{(1)}_{max} \approx 0.75   {\rm \ \ for \ \ } \rho = 1 \ \ \ . \ \ \
 \ee

\begin{figure}
\centering
\epsfysize=20cm
\mbox{\epsffile{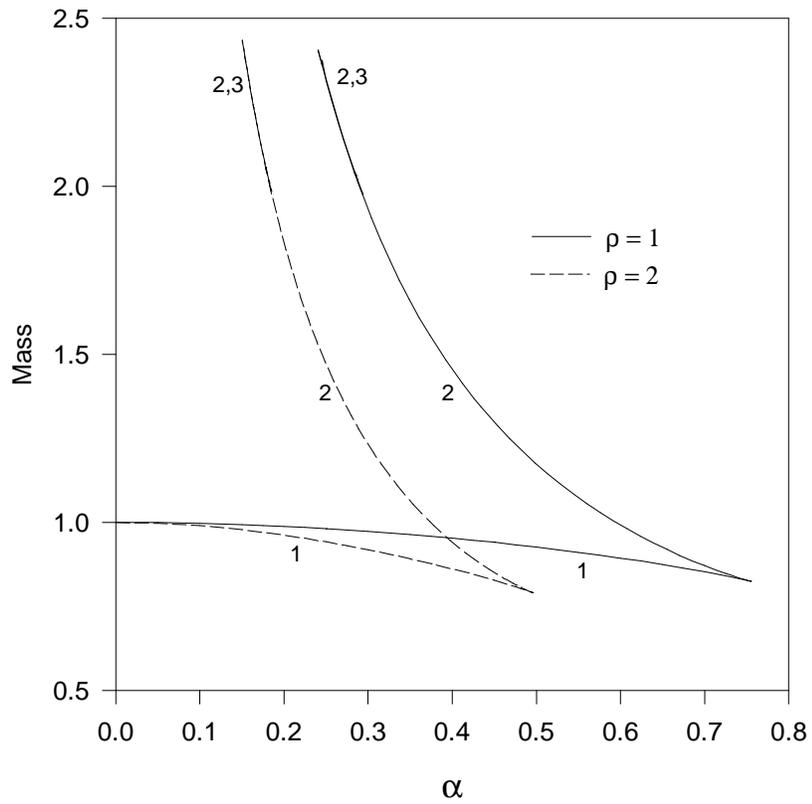}}
\caption{\label{fig2} The mass $M$ of the solutions is shown
as function of $\alpha$ for $\rho=1$ and $\rho=2$. ``1'', ``2'' and ``3'', respectively
denote the 1., 2. and 3. branch of solutions. }
\end{figure}

Our numerical analysis further 
reveals that for $\alpha \leq \alpha^{(1)}_{max}$  another branch
of solution exists. 
The mass of the solutions on this second branch is higher
than the mass of the corresponding solutions on
the first branch. This is demonstrated in Fig.~\ref{fig2}.

Our numerical results further suggest as illustrated by the figures
that the solutions on the 
second branch stop to exist at some local minimal value 
$\alpha = \alpha^{(1)}_{min}$.

To be more precise, the further evolution of the 
branches is involved~: indeed several small branches 
exist for $\alpha$ $\epsilon$ $[\alpha_{min}^{(k)}, \alpha_{max}^{(k+1)}]$
with a smaller and smaller extend in $\alpha$.
Nevertheless, it appears clearly from the figures 
that for $\alpha \rightarrow \alpha_{cr}$, where $\alpha_{cr} : = 
\alpha^{(k)}_{min}=\alpha^{(k+1)}_{max}$ for some $k$,
the value $A(0)$ tends to zero, while the values
$\vert \psi_{1,2}(0) \vert$
increase considerably and likely become infinite.
Note, however, that the additional branches are not visible in the mass plot
(see Fig.\ref{fig2}), because the numerical values are very close to
those of the second branch.
The existence of several branches was also noticed for the $n=1$ system in \cite{volkov}
and in another Einstein-Yang-Mills model in 5 dimensions \cite{bcht}.

We find numericallly:
\begin{equation}
\alpha^{(1)}_{min}\approx 0.240 \ \ , \ \ \alpha^{(2)}_{max}\approx 0.294 
\ \ , \ \ \alpha_{cr}\approx 
0.285  
\end{equation}

At the same time, the value $N_m$ stays strictly positive and tends
to the value given by (\ref{ads}) in the critical limit.
We find numerically that $N_m\approx 0.065$ for $\alpha\approx \alpha_{cr}$, 
while $c_1\approx 0.064$ for $n=2$. 
The function $N(x)$ 
reaches this minimum at a value of $x$, which tends to zero for 
$\alpha \rightarrow \alpha_{cr}$. At the same time, the function $N(x)$
becomes nearly flat on a plateau surrounding this minimum. 

The function $K(x)$ starts to develop oscillations around the
value $\sqrt{q}=0.294$. Processing on the branches, the number of oscillations
increases and becomes infinite in the critical limit. 

Very similar to what was observed in \cite{weinberg, volkov}, the
Yang-Mills domain walls get disconnected from the outside world
by an infinitely long throat in the strong gravity limit.

\subsubsection{$\rho=2$}

In order to understand the influence of two Higgs fields with different expectation
values, i.e. $\rho\neq 1$, on the domain of existence 
of the solutions, we studied in detail the case $\rho=2$.

\begin{figure}
\centering
\epsfysize=20cm
\mbox{\epsffile{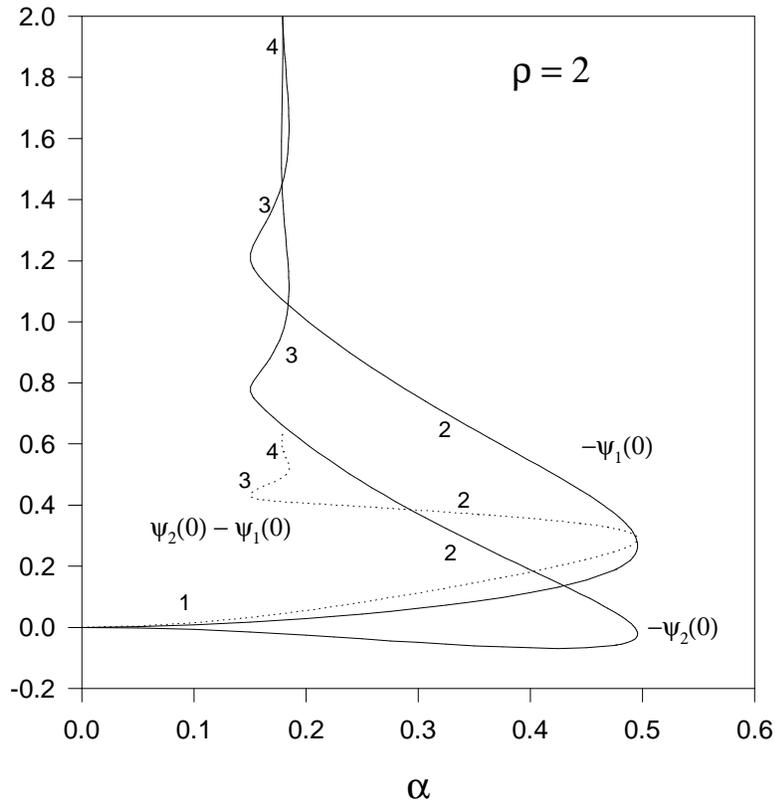}}
\caption{\label{fig3} The values of $-\psi_1(0)$ and $-\psi_2(0)$ as
 well as the difference $\psi_2(0)-\psi_1(0)$ are shown
as functions of $\alpha$ for $\rho=2$. ``1'', ``2'', ``3'' and ``4'', respectively
denote the 1., 2., 3. and 4. branch of solutions.}
\end{figure}

Varying $\alpha$, the existence of several branches of solutions is 
qualitatively very similar to the case $\rho=1$. The numerical analysis, however, 
reveals that for $\rho > 1$, the solutions exist on a smaller interval
of the coupling constant $\alpha$ in comparison to the case $\rho=1$.
Namely, we find:
\begin{equation}
\alpha^{(1)}_{max}\approx 0.496 \ , \ \alpha^{(1)}_{min}\approx 0.150 \ , 
\  \alpha^{(2)}_{max}=0.185 \ , \
 \alpha_{cr}\approx 0.179 \ \ \
{\rm for} \ \ \rho=2 \ . 
\end{equation}
In the critical limit for $\alpha\approx \alpha_{cr}$, we find
\begin{equation}
M\approx 2.05 \ \ , \ \ N_m\approx 0.068
\end{equation}
One of the apparent differences is that the two dilaton fields are non-equal. The values 
of $\psi_1(0)$ and $\psi_2(0)$ are superposed on Fig.~\ref{fig3}.
The difference $\psi_2(0)-\psi_1(0)$ is also shown and indicates
that in the critical limit, the two dilaton profiles remain significantly
different close to the origin. 

\begin{figure}
\centering
\epsfysize=20cm
\mbox{\epsffile{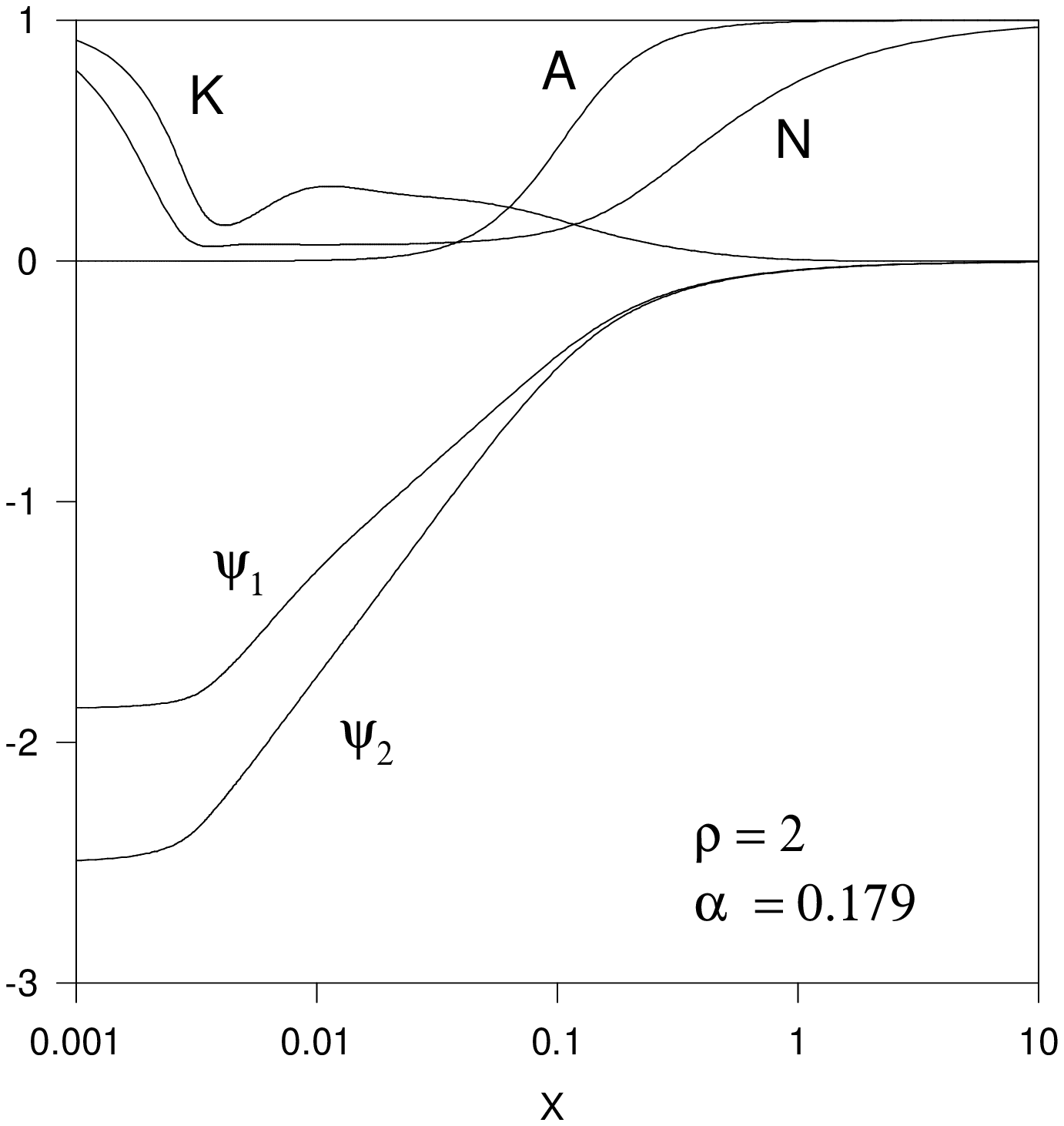}}
\caption{\label{fig4} 
The profiles of the functions  $\psi_1(x)$, $\psi_2(x)$, $K(x)$, $A(x)$ and $N(x)$  are shown
for $\rho=2$ and $\alpha$ close to the critical value $\alpha_{cr}\approx 0.179$.}
\end{figure}

The profiles of the function $K(x)$, $N(x)$, $\psi_1(x)$ and $\psi_2(x)$
are shown in Fig.~\ref{fig4} for $\alpha=0.179$. This solution is already close
to the critical solution. Note that the function $K(x)$ starts to develop
oscillations completely analogue to the case $\rho=1$.

The numerical results suggest that in the critical limit a 
solution of the analytic, non-abelian type described in the previous
subsection is reached. 
The question how this solution looks like for $\rho\neq 1$ will be addressed in
a future publication.

\section{Conclusions}
Both string theories \cite{pol} as well as
so-called ``brane worlds'' \cite{brane}, which assume the Standard
model fields to be confined on a 
3-brane (that is embedded in a higher dimensional space-time)
have enhanced the idea  that space-time
possesses more than four dimensions. In the former, the extra dimensions
are compactified
on a scale of the Planck length, while in the latter they are non-compact.
Non-perturbative, classical solutions of 
field theory models certainly play a major role in these theories.

It is therefore natural to investigate the classical solutions of
higher dimensional  Einstein-Yang-Mills theory.
In this paper, we have investigated Einstein-Yang-Mills
theories in 1+3+n dimensions, in which the matter and
metric fields are chosen to be independent of the extra $n$ coordinates.
Dimensional reduction then leads to $n$ dilaton fields
coupled individually to $n$ Higgs fields. These $n$ Higgs fields
can have independent vacuum expectation values.

The corresponding spherically symmetric equations admit several types
of solutions: (a) vacuum solutions, (b) embedded abelian Einstein-Maxwell-dilaton
solutions, (c) non-abelian solutions with diverging Higgs fields and
(d) fully non-abelian magnetically charged solutions.
The vacuum solutions (a) are trivial, while (b) and (c) were
constructed analytically for the case when all Higgs fields have
identical expectation values. The solutions of  type (d) 
had to be constructed 
numerically. 

We believe that type (b) and (c) solutions can also
be constructed analytically for generic values of the Higgs's
expectation values and we plan to reconsider this problem in a 
future publication.

It would also be interesting to 
study this model
for larger gauge groups or to adopt a different
compactification scheme of the codimension space, which here was chosen
to be $S^1 \times S^1 \times \dots S^1$. Following the investigation
for $n=1$ \cite{bbh1}, the model including a
cosmological constant is presently under investigation. 
In \cite{bbh1} the introduction of a cosmological
constant in the $(4+1)$-dimensional model led to a Liouville type potential
in the effective $4$-dimensional theory.

{\bf Acknowledgements}
Y. B. gratefully acknowledges the Belgian F.N.R.S. for financial support.
We thank E. Radu and D. H. Tchrakian for useful discussions.

 \newpage

 \newpage

 \newpage

 \newpage

\begin{thebibliography}{99}
\bibitem{foot1}
To integrate the equations, we used the differential equation
solver COLSYS which involves a Newton-Raphson method \cite{colsys}.
\end{thebibliography}
\end{document}